\documentclass[a4paper]{article}

\usepackage{icrc2013}

\title{A Heliopause Spectrum for Electrons}

\shorttitle{Heliopause Spectrum for Electrons}

\authors{M.S. Potgieter$^{1}$, R.R. Nndanganeni$^{1}$, E.E. Vos$^{1}$, M. Boezio$^{2}$ 
}

\afiliations{
$^1$Centre for Space Research, North-West University, Potchefstroom, South Africa\\ 
$^2$INFN, Sezione di Trieste, Trieste, Italy
}

\email{Marius.Potgieter@nwu.ac.za}

\abstract{We present a heliopause spectrum for galactic electrons over an energy range from 1 MeV to 50 GeV at 122 AU from the Sun. 
This spectrum can be considered the lowest possible local interstellar spectrum (LIS). 
This is obtained by using a comprehensive numerical model for solar modulation in comparison with Voyager 1 observations during 2010 and PAMELA data at Earth. 
Below $\sim$1 GeV, this LIS exhibits a power law with $E^{-(1.5\pm±0.1)}$, where $E$ is the kinetic energy of these electrons. 
Reproducing the PAMELA electron spectrum for 2009 requires a LIS with a different power law above $\sim$5 GeV of the form $E^{-(3.15\pm0.05)}$. 
The heliopause spectrum thus consists of two power laws with a break between $\sim$800 MeV and $\sim$2 GeV. }


\begin{document}
\maketitle

\section{Introduction}
A heliopause spectrum (HP) for studies on the modulation of cosmic ray (CR) electrons needs to be specified as input spectrum when using numerical models. 
This is done at an assumed modulation boundary, mostly considered to be the heliopause.  
This spectrum gets modulated throughout the heliosphere as a function of position, energy and time, a process known as the solar modulation of cosmic rays.  
The heliopause spectrum is of course related to the local interstellar spectrum (LIS), if not identical.  

The heliospheric diffusion coefficients cannot be uniquely determined so that all cosmic ray LIS at low kinetic energies $E < 5$ GeV) remain controversial because of solar modulation. 
Progress can however be made because solar modulation models have reached a relatively high level of sophistication [1,2].  	

Additionally, the Voyager 1 spacecraft is in the vicinity of the heliopause (or something that looks like it) while observing electrons between 6 and $\sim$100 MeV [3,4]. 
Together with electron observations at Earth from PAMELA with $E > 100$ MeV [5], the total modulation between the heliospheric boundary and Earth can be determined. 
This assists in providing a rather robust set of modulation parameters for a comprehensive modeling of CR modulation. 
	
Computed galactic spectra (GS) usually do not contain the contributions of any specific (local) sources within parsecs from the heliosphere so that an interstellar spectrum 
may be different from an average GS which may again be different from a LIS (thousands of AU away from the Sun), 
which might be different from a very LIS or what can be called a heliopause spectrum, right at the edge of the heliosphere, 
say within $\sim$ 200 AU away from the Sun. If known, the latter would be the ideal spectrum to use as an input spectrum for solar modulation models. 

In an attempt to make progress in this regard, the solutions of a numerical model for the modulation of galactic electrons in the heliosphere 
are presented in comparison with observations from Voyager 1 and PAMELA to obtain a computed electron spectrum at the 
heliopause that can be considered the lowest possible very LIS. This work follows on the report by Potgieter and Nndanganeni [1] on the solar modulation of electrons in the heliosphere.

\begin{figure*}[!t]
\centering
\includegraphics[width=0.8\textwidth]{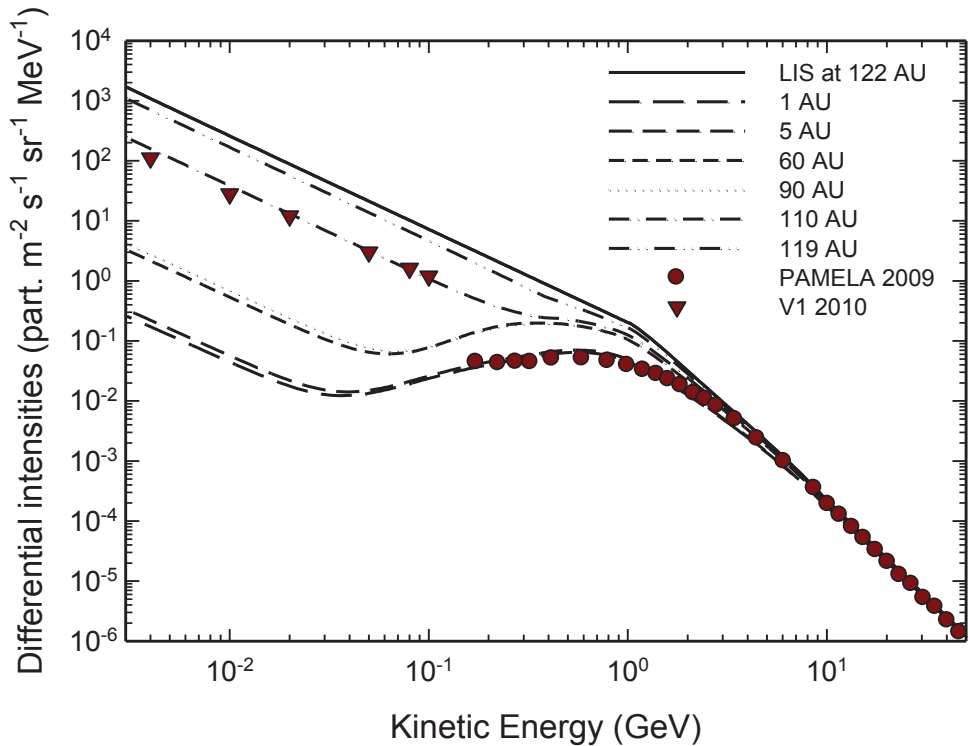}
			\caption{Computed modulated electron spectra at Earth (1 AU; with polar angle of $\theta = 90^{\circ}$), at 5 AU, 60 AU, 90 AU, 110 AU and 119 AU (with $\theta = 60^{\circ}$ corresponding to the Voyager 1 trajectory). 
			The computed spectra are compatible with observations from Voyager 1 (2010) and from PAMELA during solar minimum of 2009 as indicated. 
			The heliopause spectrum (upper solid line) is specified at the heliopause, positioned in the model at 122 AU. 
			Accordingly, the prediction is that the HP spectrum at 10 MeV may have a value of between 100 to 300 electrons 
			m$^{-2}$ s$^{-1}$ sr$^{-1}$ MeV$^{-1}$. The modulation of electrons up to Earth was discussed in detail by [1]}
\label{fig:Figure1}
\end{figure*}

\section{Numerical Model}

A full three-dimensional (3D) numerical model, with three spatial dimensions and an energy dependence (four computational dimensions), 
was used to compute electron spectra at selected positions in the heliosphere, including the inner heliosheath. 
This model was published by [1] and only changes made for this study are indicated here. In short, the model is 
based on the numerical solution of Parker’s transport equation [6] for solar modulation including all four major modulation processes. 
	
The detail on the spatial and rigidity dependence of the three major diffusion coefficients and the drift coefficient, is also not reproduced here. 
It suffices to say that in the model the heliopause is specified at 122 AU with the solar wind termination shock (TS) at 94 AU, which gives a 28 AU wide heliosheath in the 
direction in which Voyager 1 has been moving. This heliosheath plays an important role in the modulation of low energy (such as 10 MeV) galactic electrons [1,2]. 
We assume that solar modulation becomes negligible with $E > 30$ GeV; in this context, see also Potgieter and Strauss, this conference.
	
The electron spectrum from 2010 is an optimal choice from the available Voyager 1 data in order to determine a heliopause spectrum 
because earlier observed spectra, closer to the TS, are much lower and may be subjected to short-term changes as the region closer 
to the TS is more turbulent as it shifts position with changing solar activity

\section{Results and Discussion}

We use the numerical model and the available Voyager 1 observations of galactic electrons between 6 MeV to $\sim$120 MeV for 2010, when V1 moved between 112 AU and 115 AU from the Sun, to determine a heliopause spectrum. 
We began by determining the spectral slope at these energies, and then compute the differential flux values between 1 MeV and 200 MeV to reproduce the Voyager spectrum, 
at the same time addressing the amount of modulation that takes place between the modulation boundary and the observational points. 
The spectral slope was then adjusted with a different power law index above $\sim$1 GeV to reproduce the modulated PAMELA electron data at Earth [5]. 
The PAMELA electrons extend down to only $\sim$200 MeV. All The modulation parameters (including the solar wind speed and solar magnetic field) as required to reproduce the Voyager 1 and PAMELA observations are given and motivated by [1].
	
In figure 1 the computed modulated electron spectra at Earth (1 AU, with polar angle of $\theta = 90^{\circ}$), at 5 AU, 60 AU, 90 AU, 110 AU and 119 AU (with $\theta = 60^{\circ}$ corresponding to the Voyager 1 trajectory) are shown. 
The computed spectra at Earth and at 110 AU are compatible with corresponding observations from Voyager 1 and PAMELA (average for 2009) as indicated. The heliopause spectrum (upper solid line) is specified at the heliopause, positioned in the model at 122 AU. 
	
To reproduce the spectral shape of the observed Voyager spectrum, the LIS below $\sim$1.0 GeV must have a power law form with $E^{-1.5}$. Taking statistical and systematic uncertainties into account, this form can be refined to be given as $E^{-(1.5\pm0.1)}$. 
	
To reproduce the PAMELA electron spectrum for 2009, the LIS must have a power law form with $E^{-(3.15\pm0.05)}$, above $\sim$5 GeV. The latter is consistent with the spectral index reported for 40-50 GeV by the PAMELA group [5,7]. 
Combining the two power laws with a smooth transition from high to low energies produces the solid line shown in figure 1. 
This break occurs between $\sim$800 MeV and $\sim$2 GeV. We consider this spectrum as the heliopause spectrum which is the lowest possible LIS for electrons.

For diffusion coefficients independent of energy for electrons at these low energies as predicted by turbulence theory and discussed by [1], this power law for the LIS is maintained as the spectra become modulated.
Introducing any form of energy (rigidity) dependence will not give steady modulation (difference between the modulated spectrum and the LIS is unchanged) over an energy range from 1 MeV to $\sim$ 200 MeV beyond 110 AU.
	
The heliopause spectrum reported here is different (significantly higher) from the interstellar spectra reported by Webber and Higbie [8].

It should also be kept in mind that the Voyager 1 and 2 detectors cannot distinguish between electrons and positrons whereas the PAMELA detector can.
	
Evident from figure 1 is that the heliosheath acts as a very effective modulation 'hurdle' for these low energy electrons, causing intensity 
radial gradients of up to 20$\%$ AU$^{-1}$ as modelled by [2]. 
The bulk of the modulation for these electrons occurs between the heliopause and 80 AU, effectively inside the inner heliosheath. 
For an illustration of how the galactic intensity may change with decreasing distance towards the Sun, see [1,2].
	
Concerning the break in the spectral shape of the presented HP spectrum, Strong et al. [9] re-analysed synchrotron radiation data using various radio surveys to constrain 
the low-energy electron LIS in combination with data from Fermi-LAT and other experiments. They tested several propagation models based on cosmic-ray and gamma-ray data against synchrotron data from
22 MHz to 94 GHz. 
They concluded that the electron LIS must turn (exhibits a spectral break) 
below a few GeV in accord with what is presented here. Comparing their different LIS (see their figures 1, 4, 8, 11 and 12) with the one presented here, indicate that the 
LIS presented in their figure 12 resembles ours closely below $\sim$800 MeV but not above this energy. Most of their models have more-or-less the same slope above 
3 GeV but differ with ours in terms of absolute value. Their model in the top left panel of their figure 4 seems to be the most compatible. 
This means that GALPROP based model for low energies must have different galactic propagation features than the model for the higher energies.
	 
Further investigation from an astrophysics point of view seems required to establish if the $E^{-(1.5\pm0.1)}$ spectral slope below a few GeV as found here, has a galactic 
origin and as such presents a challenge to CR source models. Or, to speculate, is this power-law perhaps influenced by the solar wind TS?
It is rather curious that if the TS was considered a strong, plain shock with a compression ratio of 
$s$ = 4, the spectral shape of the consequent TS spectrum is $E^{-(1.0)}$, whereas for a weaker TS, with $s$ = 2.5, it becomes $E^{-(1.5)}$. 
The TS is much more complicated than a plain shock, so that this aspect, although unlikely, requires further study.
	
The important point is that we found, with this solar modulation approach, a break between the mentioned two power laws in the LIS for electrons at kinetic 
energies very similar to the work based on the astrophysics approach of Strong et al. [9] when utilizing synchrotron radiation to constrain the low-energy interstellar electron spectrum.
	
A final comment on CR modulation beyond the HP is that it has become a very relevant topic since both Voyager spacecraft are about to explore the 
outer heliosheath (beyond the HP) and therefore may actually measure a pristine LIS sooner than what we anticipate. Scherer at al. [10] argued that a 
certain percentage of CR modulation may occur beyond the HP. Recently, Strauss et al. [11] followed this up and computed that the differential intensity of 100 MeV protons may decrease by $\sim 25\%$
from where the heliosphere is turbulently disturbed (inwards from the heliospheric bow wave) up to the HP. However, because the diffusion coefficients 
of low energy electrons are independent of energy, making them significantly larger than for protons of the same rigidity, this percentage should be 
much less for 100 MeV electrons. Our computation for the HP spectrum at 122 AU from the Sun takes this into account. However, it could be that 
we underestimate this effect and that the true electron LIS is somewhat higher than the HP spectrum that we have presented here. 
It appears from counting rates (not differential flux) reported by [4] that Voyager 1 has observed a significant increase at what appears to be the 
heliopause. Future Voyager 1 observations should enlighten us in this respect as it moves outwards at 3 AU per year. 
By the end of 2013, Voyager 1 will be 126.4 AU away from the Sun while Voyager 2 will be at 103.6 AU, both moving into the nose region of the heliosphere
but at a totally different heliolatitude (http://voyager.gsfc.nasa.gov/heliopause/data.html).

An interesting feature occurs in the PAMELA spectra between $\sim$2 GeV and $\sim$20 GeV which may alter the heliopause 
spectrum that we present here over this energy range and is discussed in an accompaning paper, see [12].
 
\section{Conclusions}

We present a heliopause spectrum for cosmic ray electrons over an energy range from 1 MeV to 50 GeV that can be considered the lowest possible LIS. 
This is done by using a comprehensive numerical model for solar modulation in comparison with Voyager 1 observations during 2010 and PAMELA 
data at Earth. 

We report that below $\sim$1.0 GeV this LIS has a power law form with  $E^{-(1.5\pm0.1)}$ but to reproduce the PAMELA electron 
spectrum for 2009, the LIS must have a different power law form with $E^{-(3.15\pm0.05)}$ above $\sim$5 GeV. 
The latter is consistent with the spectral index reported for 40-50 GeV by the PAMELA group [5,7]. 

Combining the two power laws with a smooth transition from low to high energies produces the LIS shown as the solid line in figure 1. 
The reported break between $\sim$800 MeV and $\sim$2 GeV in the spectral form of the LIS is consistent with the results of Strong et al. [9] 
who studied different GALPROP based models for the electron LIS, from an astrophysics point of view, utilizing synchrotron radiation to constrain the low-energy interstellar electron spectrum,.

The prediction is that the electron HP spectrum at 10 MeV and at 122 AU from the Sun may have a value between 100 to 300 electrons m$^{-2}$ s$^{-1}$ sr$^{-1}$ MeV$^{-1}$, which we consider 
as the lowest possible value of the LIS at this energy.

The important point is that we found, using a solar modulation model and Voyager 1 observations in the outer heliosphere, a break between the mentioned two power laws in the LIS for
electrons.

\vspace*{0.5cm}
\footnotesize{{\bf Acknowledgment:}{~The authors thank Bill Webber for providing them with Voyager 1 electron spectrum as shown in figure 1. 
The partial financial support of the South African National Research Foundation (NRF), 
the SA High Performance Computing Centre (CHPC) and the SA Space Agency's (SANSA) Space Science Division is acknowledged.}}

\end{document}